\documentclass[prb,twocolumn,longbibliography]{revtex4-1}
\usepackage{graphicx}
\usepackage{bm}
\usepackage{wrapfig}
\usepackage{color}
\usepackage{natbib}
\usepackage{hyperref}
\usepackage{amssymb}
\usepackage{amsmath}
\usepackage{geometry}
\usepackage{comment}

\begin{document}
\title{Landauer Formula for a Superconducting Quantum Point Contact}

\author{Sergey S. Pershoguba, Thomas Veness, and Leonid I. Glazman}

\affiliation{Department of Physics, Yale University, New Haven, CT 06520, USA}

\date{\today}

\begin{abstract}
We generalize the Landauer formula to describe the dissipative electron transport through a superconducting point contact. The finite-temperature, linear-in-bias, dissipative dc conductance is expressed in terms of the phase- and energy-dependent scattering matrix of the Bogoliubov quasiparticles in the quantum point contact. The derived formula is also applicable to hybrid superconducting-normal structures and normal contacts, where it agrees with the known limits of Andreev reflection and normal-state conductance, respectively.
\end{abstract}
                             
\maketitle

The celebrated Landauer formula \cite{Landauer1957} relates the conductance of
a mesoscopic sample to the transmission coefficient for electrons passing
through it, and is valid for arbitrary transmission strength.
The derivation is usually approached via a scattering formalism, 
or the Kubo formula \cite{BruusFlensbergBook}
applied to an ensemble of noninteracting fermions. 
The former method relies on charge conservation; the latter requires performing
the calculation at a finite frequency $\omega$, followed by taking the limit
$\omega\to 0$ at small but fixed bias $\mathcal{V}$ in order to obtain the dc
conductance. 

In the case of a superconducting junction, both of these approaches are problematic.
The asymptotic scattering states are free-propagating
Bogoliubov quasiparticles 
with no well-defined charge, which 
precludes a
direct application of scattering theory. 
In the linear-response theory, the instantaneous current across the junction depends on
the phase difference
$\varphi$; and the phase perturbation, $2e\mathcal{V}/\hbar\omega$,
diverges in the limit $\omega\to 0$. 
This divergence is an indication of the ac Josephson effect\cite{josephson1962}, which predicts a
nondissipative current oscillating in  time with frequency
$2e\mathcal{V}/\hbar$. The non-perturbative in ${\cal V}$, dissipationless
alternating current component, however, generally coexists with a
linear-in-$\mathcal{V}$ dissipative one. 
Indeed, for the case of weak tunneling, the current at finite bias $\mathcal{V}$ and any temperature $T$ was found \cite{Larkin1967}
to the lowest order in transmission coefficient.
A linear-in-$\mathcal{V}$ expansion of the
current-voltage characteristic~\cite{Larkin1967} of a tunnel junction between two
superconductors \cite{TinkhamBook} yields a finite value of
the linear conductance\footnote{We assume here that the gaps in the
quasiparticle spectra of the two superconductors are not equal to each other.
 The equal-gap case exhibits a spurious divergence \cite{Larkin1967}, which is cured once
higher-order tunneling processes are accounted for, see
Eq.(\ref{conductance_SNS}).} at $T\neq 0$. This dissipative conductance $G(T)$
is caused by Bogoliubov quasiparticles tunneling across the junction. 

The perturbative-in-tunneling results are adequate for conventional large-area Josephson junctions, but are not applicable to point contacts having one or a few channels with high transmission coefficient. Such junctions are presently actively studied in a variety of platforms, including proximitized nanowires\cite{DelftProx}  and cold fermions\cite{Husmann1498,Husmann-PNAS2018}. The purpose of this work is to free the evaluation of $G(T)$ from the assumption of weak tunneling. Our main result, Eq~(\ref{trace_formula}), expresses $G(T)$ in terms of the quasiparticle scattering matrix. This generalization of the Landauer formula is valid for a junction between leads made of superconductors or normal conductors, in any combination. Additionally, the derived relation provides a lucid interpretation of the dissipative, so-called~\cite{barone1982} ``$\cos\varphi$'' component\cite{josephson1962} of the ac Josephson current.

\begin{figure}
\includegraphics[width=\linewidth]{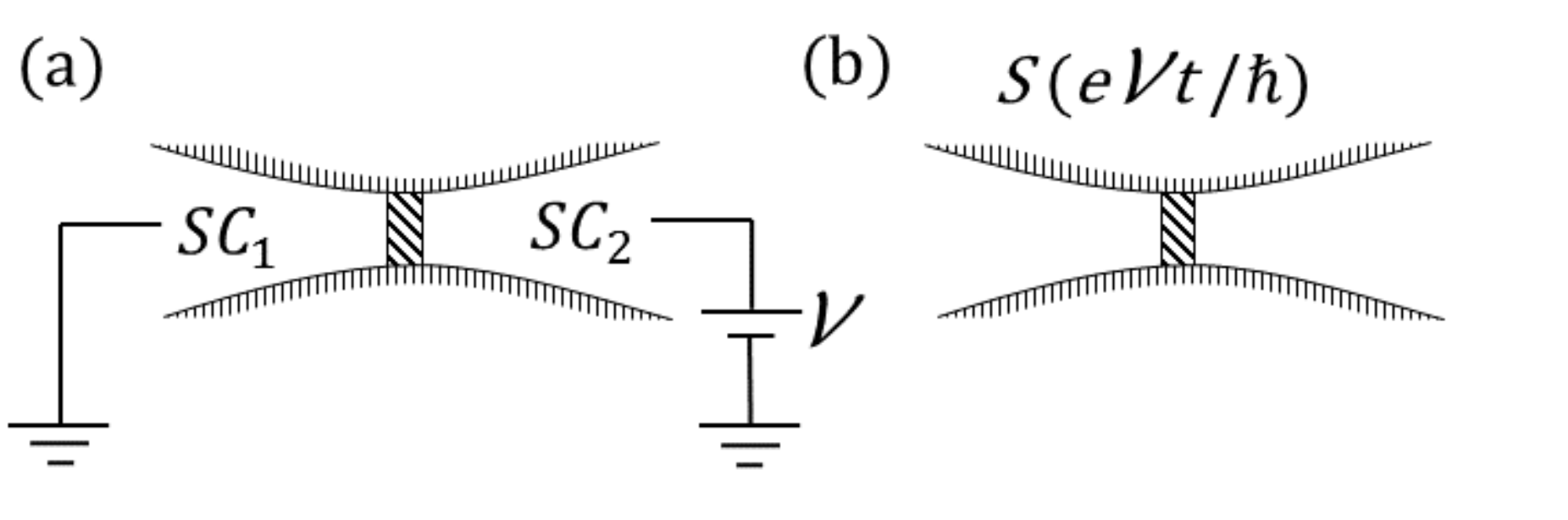}
\caption{(a) A point contact between two superconductors SC$_1$ and SC$_2$ under applied bias $\mathcal V$. (b) To evaluate the dissipative current due to the quasiparticles at finite temperature $T$, we absorb the  bias voltage in the time dependence of the quasiparticle scattering matrix $S(\Omega\,t)$, where $\Omega = e \mathcal V/\hbar$. The main general expression for dissipative conductance $G$ is given in Eq.~(\ref{trace_formula}) and application to a specific model of a superconducting point contact (SPC) in Eq.~(\ref{conductance_SNS}).} \label{fig:junction}
\end{figure}

Aiming at evaluation of $G(T)$ for a system with broken gauge invariance, it is useful to reformulate the problem so that the chemical potentials of the leads are not affected by the bias. This is achieved by introducing a time-dependent phase $e{\cal V}t/\hbar$ in the definition of the creation operators for electrons to which bias is applied, $\psi^\dagger\rightarrow\psi^\dagger\exp(ie{\cal V}t/\hbar)$ and thus endowing the scattering matrix describing the contact with a periodic dependence on time, see Fig.~\ref{fig:junction}. The time dependence allows for energy absorption by electrons passing through the junction, {\sl i.e.}, introduces channels of inelastic scattering. The energy transfer is quantized in units of $\hbar\Omega=e{\cal V}$, small in the limit ${\cal V}\to 0$.  Our strategy consists of two steps. First, we relate the scattering matrix for such ``soft'' inelastic processes to the conventionally-defined elastic scattering matrix of the system in the absence of time dependence. Next, we evaluate the absorbed power
${\cal P}$ in terms of scattering matrix and find $G(T)$ from the relation ${\cal P}=G{\cal V}^2$ for Ohmic losses. This method avoids problems associated with the charge nonconservation and presence of large nondissipative currents.
The result, Eq.~(\ref{trace_formula}), is applicable to superconducting
and hybrid normal metal--superconductor structures. For such structures,
Eq.~(\ref{trace_formula}) has the same status as that of the standard Landauer
formula for the normal-state contacts; in the absence of superconductivity,
Eq.~(\ref{trace_formula}) readily reduces to the conventional form of the
Landauer formula.

Inelastic quasiparticle scattering in channel $N$ is associated with absorption of $N$ quanta ($N=0,\pm
1, \pm 2,\dots$) and is characterized by scattering matrix $S_N$.
In order to relate $S_N$ to the elastic scattering matrix, we consider a
generic scattering problem with a Hamiltonian
\begin{align}
   H &= H_0 + W(t), \label{td_ham} \\
   W(t) &= V e^{-i\phi(t)} + V^\dagger e^{i\phi(t)}+V_0,
\nonumber
\end{align}
where $H_0$ describes the two leads, and $W(t)$ represents the coupling
between them ($V$ and $V^\dagger$ terms)
and backscattering off the junction (term $V_0$) \footnote{We added the backscattering term $V_0$ for greater generality. Its role may be illustrated by considering the contact  as a scatterer. Within the Born approximation, terms $V$ and $V^\dagger$ result in the electron transfer between the leads, while $V_0$ causes an intra-lead backscattering.}. In the case of the
time-independent phase, $\phi(t)=\phi$, scattering is elastic and described
by an instantaneous scattering matrix $S(\phi)$. At a finite bias, the phase
$\phi (t)=\Omega\, t$ winds with frequency $\Omega$, allowing for inelastic
transitions with energy transfer $N\hbar\Omega$. 

To relate $S_N$ to $S(\phi)$, we compare their respective representations by
infinite-order series in $W$. For that, we inspect the time evolution of the
wave function $|{\psi(t)} \rangle = U(t)|{m}\rangle $  with the initial state
$| m\rangle$ at $t = -\infty$; here $| m\rangle$ is an eigenstate of $H_0$ with
energy $\varepsilon_m$. The evolution operator is given by the usual
time-ordered exponential $U(t) = \mathcal T
\exp\left[\frac{1}{i\,\hbar}\int_{-\infty}^t dt_1 W_I(t_1)\right]$, and the
subscript $I$ stands for the interaction representation. The $k$-th order
expansion term of the evolution operator~\cite{SitenkoBook} reads
\begin{equation}
U_k(t)=\frac{1}{(i\,\hbar)^k}\int_{-\infty}^{t}\!\!\! dt_kW_I(t_k)\,\dots\int^{t_{2}}_{-\infty}\!\!\! dt_1 W_I(t_1).
\nonumber
\end{equation}
At this point, it is convenient to introduce a variable $s$ taking values $0,\pm 1$ and rewrite $W(t) = \sum_{s} V^{s}e^{is\phi(t)}$, where $V^{-1} \equiv V$, $V^{+1} \equiv V^\dagger$, and $V^{0}\equiv V_0$. That allows one to further specify the form of the expansion term. For $\phi(t)=\Omega\, t$, we may write $U_k(t)$ as a sum of harmonics,
\begin{eqnarray}
&& U_k(t)=\sum_N\frac{e^{iN\Omega t}}{(i\,\hbar)^k}\sum_{s_1,\dots s_k} \delta_{\sigma_k,N} \int_{-\infty}^t\!\!\!dt_k e^{i\sigma_k\Omega (t_k-t)}V^{s_k}_I(t_k)
 \nonumber\\
&&
\qquad\qquad\qquad\qquad \dots\int_{-\infty}^{t_2} d t_1e^{i\sigma_1\Omega (t_1-t_2)}V^{s_1}_I(t_1), \label{texpn}
\end{eqnarray}
with $\sigma_k = s_{k} + \ldots + s_{1}$. A similar result for the static
problem, $\phi (t)=\phi$, is obtained from Eq.~(\ref{texpn}) by replacing
 the factor $e^{iN\Omega t}\to e^{iN\phi}$ and setting $\Omega=0$ in all
the integrands. 

This form of $U_k(t)$ allows a direct comparison of the perturbative expansion of
the wave functions for linearly winding phase $\phi(t)=\Omega\,t$, and for fixed
phase $\phi(t) = \phi$, which we denote  $|{\tilde\psi(t)}\rangle$ and $|
\psi(t) \rangle$, respectively. 
Projecting the two wave functions onto the energy eigenstate $| n\rangle$ of
$H_0$ with energy $\varepsilon_n$, we find
\begin{align}
  & \langle n| \tilde \psi(t) \rangle  \equiv \langle n|\left[U(t)|_{\phi(t)=\Omega t}\right]|m\rangle  \label{td_psi} \\
  & = \delta_{nm} + \sum_N\frac{1}{i\,\hbar}\int_{-\infty}^t dt'\, e^{i(\varepsilon_{n,m}+\hbar\Omega N)t'/\hbar- 0|t'|} \, \mathcal{\tilde T}_{nm}(N,\Omega) \nonumber
\end{align}
and
\begin{align}
& \langle n| \psi(t) \rangle  \equiv  \langle n|\left[U(t)|_{\phi(t)=\phi}\right]|m\rangle\label{tid_psi}  \\
& = \delta_{nm}+\frac{1}{i\,\hbar}\int_{-\infty}^t dt'\, e^{i\varepsilon_{n,m}t'/\hbar- 0|t'|} \, \mathcal T_{nm}(\phi)\,.  \nonumber
\end{align}
The ${\cal T}$-matrices introduced above are given by the following series:
\begin{align}
& \mathcal {\tilde T}_{nm}(N,\Omega) =  \sum_{k=1}^\infty\sum_{\substack{m_{k-1},\ldots,m_{1} \\ s_k+...+s_1 = N}} \label{td_Tmatrix}\\ & \qquad\times \frac{V^{s_k}_{nm_{k-1}}\ldots V^{s_1}_{m_{1}m}}{(\varepsilon_{m,m_{k-1}}-\hbar\Omega\, \sigma_{k-1}+i0)\ldots (\varepsilon_{m,m_{1}}-\hbar\Omega\, \sigma_{1}+i0)} \nonumber
\end{align}
and
\begin{align}
& \mathcal T_{nm}(\phi) =   \label{tid_Tmatrix}
\\& \sum_N e^{i\phi N}\sum_{k=1}^\infty\sum_{\substack{m_{k-1},\ldots,m_{1} \\ s_k+...+s_1 = N}}\frac{V^{s_k}_{nm_{k-1}}\ldots V^{s_1}_{m_{1}m}}{(\varepsilon_{m,m_{k-1}}+i0)\ldots (\varepsilon_{m,m_{1}}+i0)}. \nonumber
\end{align}
Here, we introduced the notation $\varepsilon_{m,n} = \varepsilon_m-
\varepsilon_n$ and wrote the matrix elements as $V^s_{mn} = \langle m| V^{s}
| n \rangle$. 
A finite $\Omega$ brings about inelastic transitions with an arbitrary
integer number $N$ of energy quanta $\hbar \Omega$ being released ($N>0$) or
absorbed ($N<0$). The corresponding transition amplitudes are given by
$\mathcal {\tilde T}_{nm}(N,\Omega)$. In the case of fixed-phase,
$\phi(t)=\phi$, the scattering is elastic. 

By comparing the {\it inelastic} (\ref{td_Tmatrix}) and {\it elastic}
(\ref{tid_Tmatrix}) ${\cal T}$-matrices, we note that in the limit $\Omega \to
0$
\begin{align}
    \mathcal{\tilde T}_{nm}(N,0) = \int_{0}^{2\pi}\,\frac{d\phi}{2\pi}\, \mathcal T_{nm}(\phi)\,e^{-i\phi N}. \label{elastic_inelastic_relation}
\end{align}
 The utility of this expression is that the scattering matrix of a
time-independent problem may be easier to evaluate. The use of
Eq.~(\ref{elastic_inelastic_relation}) is justified as long as the effect of
$\hbar\Omega$ in the energy denominators of Eq.~(\ref{td_Tmatrix}) is
negligible. An applicability criterion specific to a superconducting junction
is discussed at the end of the Letter. We note in passing that
Eq.~(\ref{elastic_inelastic_relation}) agrees with the ``frozen scattering
matrix'' principle set forward in Refs.~[\onlinecite{MoskaletsButtiker},\onlinecite{Arrachea2006}]. 
Next, we evaluate dissipative conductance using Eq.~(\ref{elastic_inelastic_relation}).

The dissipated power may be written using scattering theory, where the absorbed
power, averaged over states in equilibrium, is \footnote{In writing the expression for power, we assume that the fermionic states are double-degenerate due to spin. The corresponding factor of $2$ cancels with the factor $1/2$, which corrects for the double-counting over the indices $n,m$ in the expression for power~(\ref{power}).}
\begin{align}
    & \mathcal P = \frac{2\pi}{\hbar} \sum_{N} N\hbar \Omega \sum_{n,m}\left| \mathcal{\tilde T}_{nm}(N,\Omega) \right|^2 \label{power}  \\
    & \quad \times \left[f(\varepsilon_n)-f(\varepsilon_m)\right]\, \delta(\varepsilon_n-\varepsilon_m+\hbar\Omega N). \nonumber
\end{align}
Each term in the sum over $N$ here has a simple meaning: it is a product of the
energy $N\hbar\Omega$ absorbed in a transition, multiplied by the transition
rate (here $f(\varepsilon_{n,m})$ are fermionic occupation factors). 
In the framework of scattering theory, it is customary to work in the
continuous energy representation instead of the discrete indices $n$ and $m$.
Therefore, we replace $n \to (\varepsilon'\alpha)$, $m \to (\varepsilon\beta)$
and introduce the  density of states $\rho_\alpha(\varepsilon')$ and
$\rho_\beta(\varepsilon)$ to re-write  Eq.~(\ref{power}) in the form
\begin{align}
    \mathcal P &= \frac{2\pi \hbar\Omega}{\hbar} \sum_{N}  N\,\sum_{\alpha,\beta}\iint d\varepsilon'd\varepsilon\, \rho_\alpha(\varepsilon') \rho_\beta(\varepsilon) \label{power1}\\
    & \times\left|\mathcal{\tilde T}_{\varepsilon'\alpha \,,\,\varepsilon\beta}(N,\Omega)\right|^2  \,\left[f(\varepsilon')-f(\varepsilon)\right] \,\delta(\varepsilon'-\varepsilon+\hbar\Omega N). \nonumber
\end{align}    
Here, $\alpha$ and $\beta$ are the residual discrete indices; they may label
channels, leads, particle-hole branches, etc. We integrate Eq.~(\ref{power1})
over $\varepsilon'$ and expand to the lowest (second) order in $\Omega$
\begin{align}
   \mathcal P &= \frac{2\pi (\hbar\Omega)^2}{\hbar} \sum_{N} N^2\,\sum_{\alpha\beta}\int d\varepsilon\, \rho_\alpha(\varepsilon)\rho_\beta(\varepsilon) \label{power2} \\ & \qquad\qquad\times\left| \mathcal{\tilde T}_{\varepsilon\alpha \,,\,\varepsilon\beta}(N,0)\right|^2 \,[-\partial_{\varepsilon}f(\varepsilon)]. \nonumber 
\end{align}
Crucially, the {\it inelastic} ${\cal T}$-matrix $\tilde {\cal T}(N,\Omega =
0)$ is evaluated at $\Omega = 0$ in Eq.~(\ref{power2}). So we may express it
via the {\it elastic} ${\cal T}$-matrix according to
Eq.~(\ref{elastic_inelastic_relation}),
\begin{align}
   & \mathcal P = \frac{2\pi (\hbar\Omega)^2}{\hbar} \int d\varepsilon\, [-\partial_{\varepsilon}f(\varepsilon)] \iint\limits_{0}^{\quad2\pi} \frac{d\phi'd\phi}{(2\pi)^2} \label{power3} \\ & \times \sum_{N} e^{i N(\phi'-\phi)}  N^2\,\sum_{\alpha\beta}\rho_\alpha(\varepsilon)\rho_\beta(\varepsilon)\, {\mathcal T}^\ast_{\varepsilon\alpha \,,\,\varepsilon\beta}(\phi')\mathcal T_{\varepsilon\alpha \,,\,\varepsilon\beta}(\phi). \nonumber 
\end{align}
Next we use the relation\cite{SitenkoBook},\footnote{We follow Ref.~[\onlinecite{SitenkoBook}], where the relation is derived  $\mathcal S_{nm} = \delta_{nm} -2\pi i\,\delta(\varepsilon_n-\varepsilon_m) \,\mathcal T_{nm}$ between matrix elements of scattering $\mathcal S$ and transition $\mathcal T$ operators. In the energy $\varepsilon$ representation, this relation translates into the following identification of the scattering matrix $S_{\alpha\beta}(\varepsilon) =\delta_{\alpha\beta}  - 2\pi i\,   \sqrt{\rho_\alpha(\varepsilon)\rho_\beta(\varepsilon)}\,\,\mathcal T_{\alpha\varepsilon,\beta\varepsilon}$.} between the ${\cal T}$-matrix and the on-shell {\it elastic} scattering matrix and replace the derivatives $-2\pi i \sqrt{\rho_\alpha(\varepsilon)\rho_\beta(\varepsilon)}\,\partial_\phi \mathcal T_{\varepsilon\alpha \,,\,\varepsilon\beta}(\phi) \to \partial_\phi S_{\alpha\beta}(\phi,\varepsilon)$, which allows one to express the summation over $\alpha$ and $\beta$ as a trace. Further simplification comes from noticing that $\sum_{N} e^{i N(\phi'-\phi)}  N^2 = 2\pi\partial_\phi\partial_{\phi'}
\delta(\phi-\phi')$ in Eq.~(\ref{power3}).
Finally, recalling that $\Omega = \mathcal V e/\hbar$ and $G = \mathcal P/\mathcal V^2$, we obtain the dissipative conductance, which is the main result of this work:
\begin{align}
  G = \frac{e^2}{h} \int\limits d\varepsilon\, [-\partial_{\varepsilon}f(\varepsilon)] \int\limits_{0}^{2\pi} \frac{d\phi}{2\pi} \,   {\rm Tr}\left\{ \partial_{\phi}  S^\dagger(\phi,\varepsilon)\,\partial_{\phi} S(\phi,\varepsilon) \right\}. \label{trace_formula}
\end{align}
Consistently with Eq.~(\ref{td_ham}), the gauge in Eq.~(\ref{trace_formula}) is fixed by associating the phase factor $e^{i\phi}$ with the transmission amplitude of the normal-state scattering matrix. For a superconducting junction, the order parameter phase difference across the junction is $\varphi=2\phi$.

It is instructive to relate the dc conductance $G$ to the
dissipative part of the low-frequency admittance $Y(\omega\to 0,
\phi,T)$ of the same junction.~\footnote{The admittance is defined as a linear ac response to an applied bias, $I(\omega)=Y(\omega,\phi,T)U(\omega)$.}
In evaluating ${\rm Re}\,Y({\omega\to 0},\phi, T)$, the perturbation
$\delta\phi(t)=eU\cos(\omega t)/\hbar\omega$ of the phase $\phi
(t)=\phi+\delta\phi(t)$ across the junction is a small parameter, as the
limit $U\to 0$ is taken first. Applying the same technique as above, we find
that only single-quantum transitions occur to linear order in $U$, {with
amplitudes $\propto\partial_\phi S$} . Evaluation of the absorption power
yields~\footnote{{In the absence of superconductivity, Eq. (13) is obtainable
also within the formalism of emissivities \cite{Buttiker1994}}}
\begin{eqnarray}
&&{\rm Re}\,Y(\omega\to 0,\phi, T)  \label{rey}\\
&&=\frac{e^2}{h} \int\limits d\varepsilon\, [-\partial_{\varepsilon}f(\varepsilon)]\, {\rm Tr}\left\{ \partial_{\phi}  S^\dagger(\phi,\varepsilon)\,\partial_{\phi} S(\phi,\varepsilon) \right\}. \nonumber
\end{eqnarray}
Comparing Eq.~(\ref{trace_formula}) with (\ref{rey}) and recalling that the
phase winds with time as $e{\cal V}t/\hbar$, we conclude that $G$ may be viewed as a
time-averaged value
\begin{equation}
G=\overline{{\rm Re}\,Y(\omega\to 0,e{\cal V}t/\hbar, T)}
\label{Grey}
\end{equation}
of the instantaneous conductance given by the dissipative part of the admittance. It generalizes the known relation in normal junctions \cite{BruusFlensbergBook} between the dc Landauer conductance and the $\omega \to 0$ limit of the Kubo formula. 

Equation~(\ref{trace_formula}) is non-perturbative in tunneling, which is one of its advantages over the known~\cite{TinkhamBook,Larkin1967} results. 
We illustrate the utility of Eq.~(\ref{trace_formula}) by finding the conductance between two superconductors connected by a short channel of arbitrary transmission coefficient, see
Fig.~\ref{fig:junction}. 
Finite temperature induces a thermal population of quasiparticles in each of
the two leads. To start with, we focus on the case of equal gaps $\Delta_1 =
\Delta_2 = \Delta$. We follow Ref.~[\onlinecite{Beenakker1991}] and evaluate the
corresponding $S$-matrix. In the Bogoliubov-de Gennes representation, the
quasiparticle excitations have positive energy $\varepsilon>\Delta$, and the
S-matrix is 4-by-4 due to the 2 leads and 2 particle-hole branches, see
~[\onlinecite{* [{See Appendix}] [{ for further details.}] SM}] 
for details. We apply Eq.~(\ref{trace_formula}) and
evaluate the conductance at arbitrary transmission coefficient $\tau$ of the junction,
\begin{align}
    \frac{G_{SPC}}{G_n} = \int_\Delta^\infty d\varepsilon \,[-\partial_\varepsilon f(\varepsilon)]\, \frac{2\,\varepsilon^2}{\sqrt{(\varepsilon^2-\Delta^2)(\varepsilon^2-\Delta^2(1-\tau))}}. \label{conductance_SNS}
\end{align}
Here $G_n = 2e^2\tau/h$ is the normal-state conductance. An alternative
way to derive Eq.~(\ref{conductance_SNS}) is to use Eq.~(\ref{Grey}) and the
result \cite{Kos2013} for ${\rm Re}\,Y(\Omega,\phi,T)$.

\begin{figure}
\includegraphics[width=0.9\linewidth]{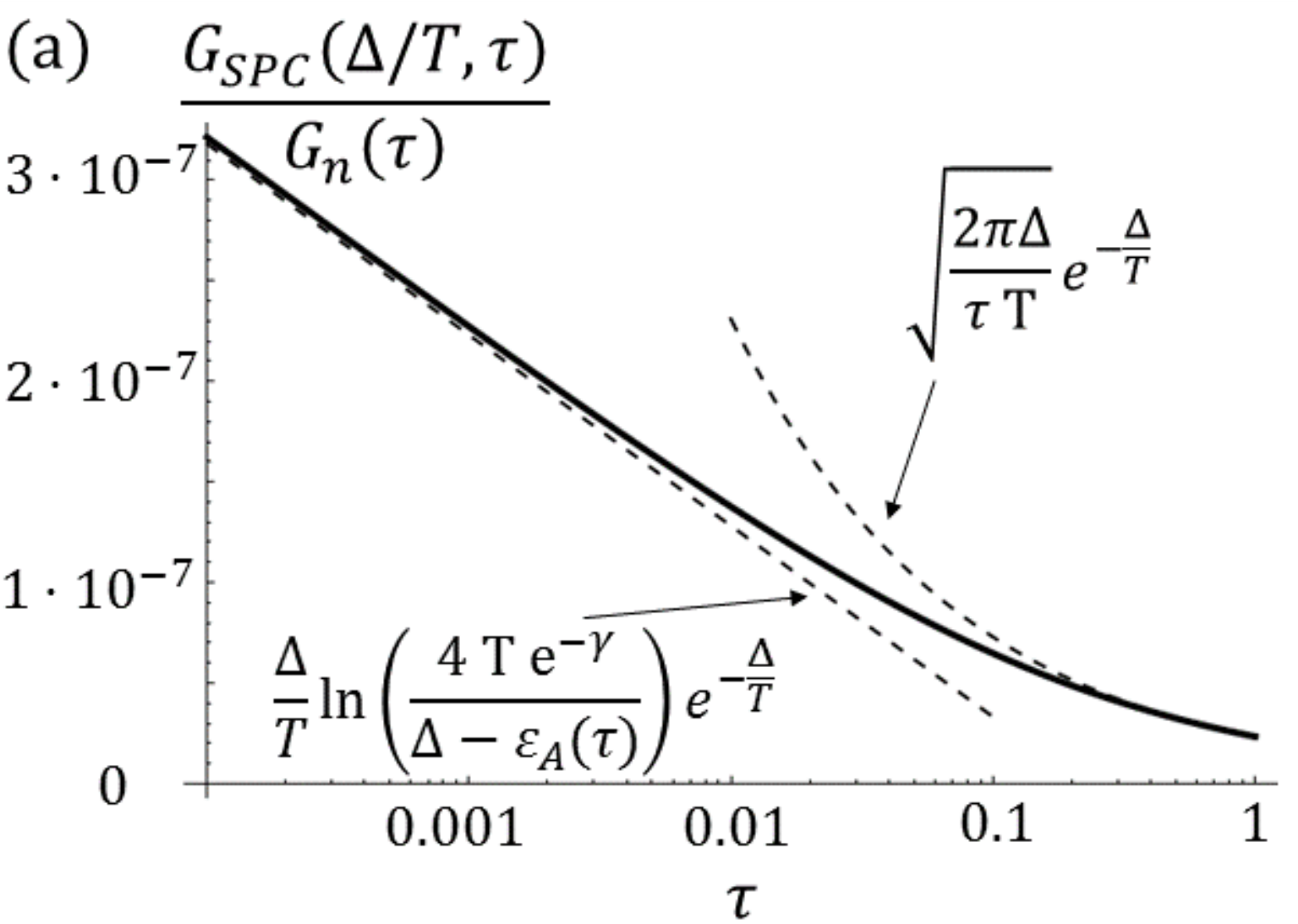}  \\
\includegraphics[width=0.9\linewidth]{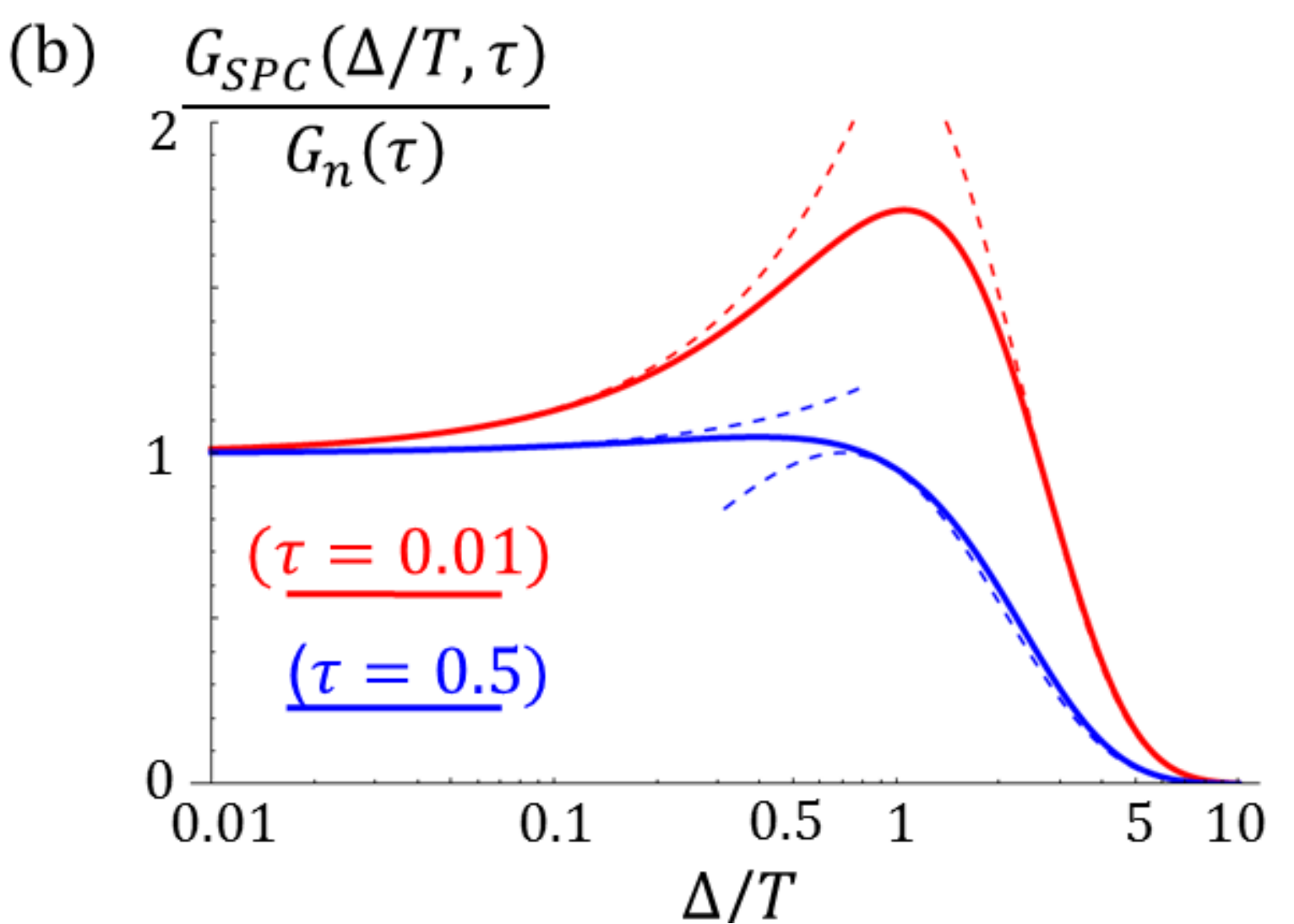}  \caption{(a) Conductance $G_{SPC}$ of a superconducting point contact as a function of transmission coefficient $\tau$ evaluated from Eq.~(\ref{conductance_SNS}) at a low temperature, $T/\Delta = 0.05$. The two asymptotes
of Eq.~(\ref{large_gap_asymptote}), shown in dashed lines, are valid, respectively, at transmission $\tau \ll T/
\Delta$ and $\tau \gg T/\Delta$. 
(b) $G_{SPC}$ as a function of $\Delta/T$ (solid lines) at two fixed values of $\tau$, along
 with the asymptotes~(\ref{large_gap_asymptote}) and (\ref{small_gap_asymptote}), shown by dashed lines. } 
\label{fig:conductanceSNS}
\end{figure}

It is instructive to consider first the low-temperature asymptote, $\Delta/T
\gg 1$,
\begin{align}
    &\frac{G_{SPC}(\Delta/T,\tau)}{G_n(\tau)} \label{large_gap_asymptote} \\
    & \qquad\approx \sqrt{\frac{2\Delta}{\Delta+\varepsilon_A(\tau)}}\, \frac{\Delta}{T}\, e^{-\frac{\Delta+\varepsilon_A(\tau)}{2\,T}} K_0\left[\frac{\Delta-\varepsilon_A(\tau)}{2T}\right], \nonumber
\end{align}    
where $K_0(x)$ is the modified Bessel function. Note that the superconducting
contact supports  Andreev levels with energies $\varepsilon_A(\tau,\phi) =
\Delta \sqrt{1-\tau \sin^2\phi}$ carrying the Josephson current, which is not
the subject of this work. However the indirect effect of the Andreev levels is
observed in Eqs.~(\ref{conductance_SNS}) and (\ref{large_gap_asymptote}), where
we denote $\varepsilon_A(\tau) \equiv \varepsilon_A(\tau,\pi/2) =
\Delta\sqrt{1-\tau}$. The Andreev levels lead to a strong modification of the
density of states of the delocalized quasiparticles and thus influence their
transport. 
The low-temperature conductance~(\ref{large_gap_asymptote}) displays a crossover
between two asymptotes defined by a dimensionless ratio 
$\frac{\Delta-\varepsilon_A}{T} \propto \frac{\tau\Delta}{T}$. Above the crossover temperature ($T\gg \tau\Delta$), the conductance may be approximated as $G_{SPC} = G_n\frac{\Delta}{T} 
\ln[4\,e^{-\gamma}\, T/(\Delta-\varepsilon_A)]e^{-\frac{\Delta}{T}}$ (here $\gamma$
is the Euler-Mascheroni constant). 
We note that the perturbative-in-$\tau$
result\cite{Larkin1967,TinkhamBook} which diverges as
$2e{\cal V}\to 0$, is cut off by the scale $\Delta-\varepsilon_A$.
Below the crossover temperature ($T \ll\tau\Delta$), one
finds $G_{SPC} = G_n \sqrt{{\frac{2\pi\Delta}{\tau T}}}e^{-\frac{\Delta}{T}}$.
Both asymptotes are illustrated in Fig.~\ref{fig:conductanceSNS}(a). 

The high-temperature $T \gg \Delta$ (i.e. small-gap) asymptote is
\begin{align}
    \frac{G_{SPC}(\Delta/T,\tau)}{G_n(\tau)} \approx  1 + \frac{\Delta}{T}\,k(\tau)\,.  \label{small_gap_asymptote}
\end{align}
Note that the coefficient $k(\tau)\geq 0$ (see Ref.~[\onlinecite{SM}] for the full expression). It is logarithmically large, $k(\tau) \sim -\frac{1}{4}
\ln \tau$, for $\tau \rightarrow 0$, and $k(\tau=1) = 0$ .  Therefore, at any
$\tau<1$ the conductance $G_{SPC}$ initially grows with the opening of the
superconducting gap $\Delta$. We plot the dependence of $G_{SPC}$ on $\Delta/T$
in Fig.~\ref{fig:conductanceSNS}(b) and observe that the conductance reaches
maximum at $\Delta \sim T$. Note that the thermoelectric transport coefficients
of SPC exhibit similar behavior\cite{thermopower2019}.

The dissipative conductance Eq.~(\ref{conductance_SNS}) involves an unusual type of multiple Andreev reflection processes. In such events,
quasiparticles are not created but rather gain energy exceeding $e{\cal V}$ at
$N>1$.
In the context of Eqs.~(\ref{power})--(\ref{power2}), $N$ represents the number
of energy quanta $\hbar \Omega$ absorbed or emitted during the quasiparticle
tunneling. Because of the relation $\hbar \Omega = e \mathcal V$, integer $N$
also has the meaning of the number of electrons passing through the junction in
a scattering event. The corresponding probabilities are
given by the appropriately thermally-averaged\cite{SM} values of $|\mathcal{\tilde
T}(N,\Omega)|^2$, see Eq.~(\ref{power2}). {At $T\ll\Delta\cdot\tau$, the averaged $|\mathcal{\tilde T}(N,\Omega)|^2$ depend weakly on $N$ for
$N < N^\ast = \sqrt{\Delta \tau/T}$ and decay as $\left|\tilde{T}(N,\Omega)\right|^2 \sim
1/N^4$ for $N> N^\ast$. This indicates that processes with a transfer of a
large number of electrons gain significance at low temperatures.
} 

If both leads are superconducting, the series for the absorbed power
(\ref{power2}) contains infinitely many terms in $N$, and the trace formula
(\ref{trace_formula}) is an agile way to calculate $G$. If at least one of the
leads is non-superconducting, the sum over $N$ in Eq.~(\ref{power2}) truncates.
As an example, we consider an NS junction, i.e. set $\Delta_1 = 0$, $\Delta_2 =
\Delta$. It is easy to see~\cite{SM} that the highest harmonics of the elastic S-matrix are $e^{\pm 2i\phi}$, truncating the series at $|N|=2$. Evaluating the sum or using
the trace formula (\ref{trace_formula}), and accounting for the unitarity of the
S-matrix, we recover the known~\cite{BTK1982} expression,
\begin{eqnarray}
    G_{NS} = \frac{2e^2}{h} \int_0^{\infty} d\varepsilon\,[&-&\partial_\varepsilon f(\varepsilon)]\left[(1 -|r^{ee}|^2+|r^{he}|^2)\right.
    \nonumber\\
    &+&(1-|r^{hh}|^2+\left.|r^{eh}|^2)\right]\,,
    \label{GNS}
\end{eqnarray}
where $r^{ee}(\varepsilon)$, $r^{hh}(\varepsilon)$, and $r^{he}(\varepsilon)$,
$r^{eh}(\varepsilon)$  are, respectively, the particle, hole, and two
Andreev reflection amplitudes \footnote{Equation~(\ref{GNS}) does not assume any specific model of the scatterer, while Ref.~[\onlinecite{BTK1982}] considers a concrete ``delta-function'' scatterer model.}. The S-matrix of a normal junction
($\Delta_1=\Delta_2=0$) contains only $e^{\pm i\phi}$ harmonics, along with a
$\phi$-independent part. As a result, $r^{he}(\varepsilon)=
r^{eh}(\varepsilon)=0$ and Eq.~(\ref{GNS}) reduces to the standard Landauer
formula in the particle-hole representation.

In the derivation of Eq.~(\ref{trace_formula}), we relied upon the relation
between elastic and ``soft'' inelastic scattering matrices, cf.
Eq.~(\ref{elastic_inelastic_relation}). This is justified as long as $\hbar
\Omega$ is negligible compared to the typical energy differences $\varepsilon_m
- \varepsilon_{m'}$ involved in the summation over virtual states.
{In the context of a tunnel junction between two superconductors with gaps
$\Delta_1\neq\Delta_2$, one may estimate the significance of the next-order in
$\Omega = e \mathcal V/\hbar$ terms by expanding in ${\cal V}$ the
known~\cite{Larkin1967} expression, $I(\mathcal V) =  I_1(\mathcal V) +
I_3(\mathcal V) + \mathcal O(\mathcal V^5)$, where $I_n(\mathcal V) \propto
\mathcal V^n$. We evaluate the ratio of the consecutive terms in the expansion
of current\cite{SM} and find $\frac{I_3}{I_1} \propto \frac{(e\mathcal V)^2}{T^2}$ and
$\frac{I_3}{I_1} \propto \frac{(e\mathcal V)^2}{(\Delta_1 - \Delta_2)^2}$ in
the cases $|\Delta_1-\Delta_2| \gg T$ and $|\Delta_1-\Delta_2| \ll T$,
respectively. In other words, the next-order corrections in $e\mathcal V$ may
be dropped as long as $\hbar \Omega = e\mathcal V$ is the smallest energy scale
in the problem. At finite transmission $\tau$ and equal gaps, for which Eq.~(\ref{conductance_SNS}) is derived, this applicability criterion
amounts to $e \mathcal  V \ll {\rm min}[\,T,\Delta - \varepsilon_A(\tau)]$}.

{It is worth emphasizing that the derived dissipative conductance $G_{SPC}$,
Eq.~(\ref{conductance_SNS}), is entirely due to the itinerant Bogoliubov
quasiparticles passing through the junction. The associated Andreev levels do
not contribute to the dissipation in the absence of relaxation. The latter
creates an additional channel of dissipation via the Debye
mechanism~\cite{Debye1912}. To quantify this, we introduce a phenomenological
relaxation rate $\gamma$ for an occupied Andreev
level~\cite{AverinAdiabatic1996} and estimate the ratio $\frac{I_A}{I_{qp}}$ of
the dissipative current $I_A(\mathcal V)$ due to the
Andreev levels\cite{SM} and the current $I_{qp}(\mathcal V) = G_{SPC}\mathcal V$ due to
the quasiparticles. In the limit $\frac{\tau\Delta}{T} \ll 1$, we estimate
$\frac{I_A}{I_{qp}} \propto  \frac{\tau}{\ln(T/\tau\Delta)}\frac{\Delta\,\hbar
\gamma}{(\hbar\gamma)^2+(2e\mathcal V)^2} $, indicating that the quasiparticle
current $I_{qp}$ dominates even in the linear-in-${\cal V}$ regime ($e{\cal
V}\ll\hbar\gamma$) provided the relaxation rate $\hbar \gamma \gtrsim
\tau\Delta$. In the limit of low temperatures $T/\Delta \ll 1$ and intermediate
$\tau$, we find that the ratio of currents scales as $\frac{I_A}{I_{qp}}
\propto \frac{T}{\hbar\gamma} \exp[\frac{\Delta}{T}(1-\sqrt{1-\tau})]$ and
$\frac{I_A}{I_{qp}} \propto \frac{T \,\hbar\gamma}{(e \mathcal V)^2}
\exp[\frac{\Delta}{T}(1-\sqrt{1-\tau})]$ in the opposite regimes of small
($e\mathcal V \ll \hbar \gamma$) and large ($e\mathcal V \gg \hbar \gamma$) bias, respectively. In the
latter regime, the large exponential factor may be mitigated  by a small $\gamma$. Note that in the
absence of the relaxation due to phonons as, e.g., in the cold atom experiments
\cite{Husmann-PNAS2018}, the relaxation is itself determined by the
quasiparticle population and is, therefore, exponentially suppressed at low temperatures, $\gamma
\propto \exp(-\Delta/T)$ .}

In summary, we have expressed the dissipative linear conductance $G$ of a
superconducting quantum point contact in terms of the scattering matrix for
Bogoliubov quasiparticles, see Eq.~(\ref{trace_formula}). At a finite
temperature, $G$ is finite; Eq.~(\ref{trace_formula}) adequately accounts for
the thermally-excited quasiparticles passing through the junction. It
generalizes the Landauer formula and is valid for junctions with normal or
superconducting leads. In addition, we uncovered the relation (\ref{Grey}) between the dc conductance and the phase-averaged real part of the ac admittance of a junction. 

This work is supported by the DOE contract DE-FG02-08ER46482 (LIG), the ARO
grant W911NF-18-1-0212 (SSP), and by the Yale Prize Postdoctoral Fellowship
(TV).

\bibliography{biblio}
\appendix  
\newpage
~
\newpage
\section{Elastic scattering matrix of a superconducting point contact at arbitrary $\Delta_1/\Delta_2$\,.}

\begin{figure}
 \includegraphics[width=0.9\linewidth]{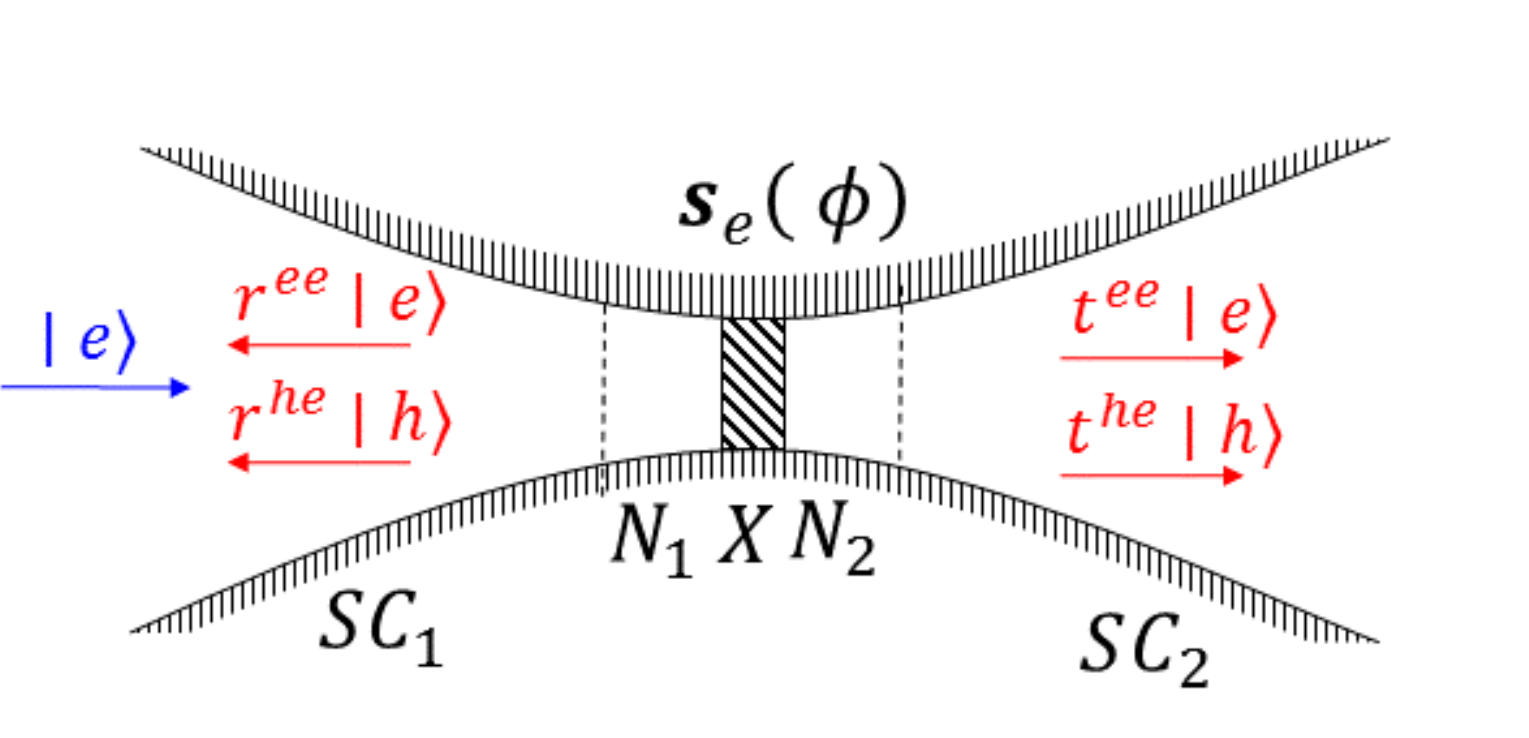}  
 \caption{Bogoliubov-de Gennes model used to derive the scattering matrix of a single-channel superconducting point contact (SPC). The normal regions $N_1$ and $N_2$ are introduced for convenience of formulating a scattering problem. A typical scattering process is demonstrated with arrows: an incident particle-like quasiparticle (emphasized in blue) from the left superconductor scatters as particle-like  or hole-like quasiparticles in both superconductors. The scattering matrix that describes such processes (\ref{S_SNS}) is 4-by-4.} 
\label{fig:SPC}
\end{figure}

We consider a single-channel quantum contact shown in Fig.~\ref{fig:SPC}. In particle-hole representation, a typical scattering process of a Bogoliubov quasiparticle is demonstrated in colored arrows. A particle-like quasiparticle incident from the left superconductor is scattered in four channels (2 particle-hole and 2 leads) with the corresponding scattering amplitudes $r^{ee},r^{he},t^{ee}$, and $t^{he}$. It is convenient to collect all scattering amplitudes in a single 4-by-4 scattering matrix  $[\psi^{-}_{e1},\psi^{+}_{e2},\psi^{-}_{h1},\psi^{+}_{h2}]^{\rm T} = \bm S(\phi,\varepsilon)\,[\psi^{+}_{e1},\psi^{-}_{e2},\psi^{+}_{h1},\psi^{-}_{h2}]^{\rm T}$ relating the incoming and outgoing states; here the superscript signs $\pm$  denote the direction of the group velocity. We follow Beenakker~\cite{Beenakker1991} and generalize the scattering matrix to the case of non-equal gaps $\Delta_1 \neq \Delta_2$
\begin{align}
   & \bm S(\phi,\varepsilon) = \bm B(\varepsilon) \left[\bm I-\bm S_n(\phi) \bm A(\varepsilon) \bm \right]^{-1}\,\left[\bm S_n(\phi)-\bm A(\varepsilon) \right]\bm B^{-1}(\varepsilon), \nonumber\\
   & \quad\bm  S_n(\phi) = \left( 
   \begin{array}{cc}
        \bm s_e(\phi) & \bm 0 \\
        \bm 0 & \bm s_h(\phi) 
   \end{array} \right),\quad \bm A(\varepsilon) = \left( 
   \begin{array}{cc}
       \bm 0  & \bm a(\varepsilon) \\
       \bm a(\varepsilon)  & \bm 0  
   \end{array} \right), \nonumber \\
   & \quad \bm B(\varepsilon) = \left( 
   \begin{array}{cc}
       \sqrt{\bm 1 - \bm a^2(\varepsilon)}  & \bm 0 \\
       \bm 0  & \sqrt{\bm 1 - \bm a^2(\varepsilon)}  
   \end{array} \right).\label{S_SNS}
\end{align}
Within the Supplement, we choose a convention in which the bold upper-case (e.g. $\bm S,\bm S_n$, $\bm A$, etc.) and lower-case (e.g. $\bm s_e,\bm s_h$, $\bm a$, etc.) letters denote the 4-by-4 and 2-by-2 matrices, respectively. The matrix $\bm S_n$ describes the scatterer X in a normal state; its diagonal blocks $\bm s_e$ and $\bm s_h=\bm s_e^\ast$ act in the particle and hole subspaces. As stated after Eq. (12) of the main text, we absorb the phase $\phi$ into the off-diagonal elements of the scattering matrix
\begin{align}
    \bm s_e(\phi) = \left(\begin{array}{cc}
    r & i\,t\,e^{-i\phi} \\ 
    i\,t\,e^{i\phi} & r
    \end{array}\right). \label{S_N}
\end{align}
These matrix elements define the transmission amplitudes. This gauge is most convenient for the generalizations involving application of a voltage bias to the junction. Focusing on a short channel, we
assume that the reflection $r$ and transmission $t$ amplitudes are energy independent. Note that the conventional Josephson phase difference $\varphi$ is related with the defined phase $\phi$ as $\varphi = 2\phi$. The matrices 
\begin{align}
    \bm a(\varepsilon) =  \left(\begin{array}{cc} 
    a_1(\varepsilon) & 0 \\ 
    0 & a_2({\varepsilon})
    \end{array}\right),\quad a_{1,2}(\varepsilon)  = \frac{\varepsilon-\sqrt{\varepsilon^2-\Delta^2_{1,2}}}{\Delta_{1,2}}, 
\label{a12}
\end{align}
and $\bm A(\varepsilon)$ describe the Andreev reflection at the opposite ends of the channel, and matrix $\bm B(\varepsilon)$ describes the transmission at its boundaries. The form of $a_{1,2}(\varepsilon)$ in Eq.~(\ref{a12}) allows for non-equal gaps $\Delta_1 \neq \Delta_2$. 

The scattering matrix~(\ref{S_SNS}) is valid at energies $\varepsilon \sim \Delta \ll E_F$.

\section{Conductance of a short channel connecting two superconductors with $\Delta_1 = \Delta_2 = \Delta$.}
Below, we provide a detailed derivation of Eq.~(15) for the dissipative conductance of a superconducting point contact (SPC), starting with
Eq.~(12) of the main text.

(i) In the case of equal gaps $\Delta_1 = \Delta_2 = \Delta$, the scattering matrix~(\ref{S_SNS}) becomes
\begin{align}
   \bm S = [\bm I - a \bm S_n \bm\tau_x]^{-1} [\bm S_n - a \bm \tau_x] \label{S_equal_gaps}
\end{align}
where $a = [\varepsilon-\sqrt{\varepsilon^2-\Delta^2}]/\Delta$, and we defined the 4-by-4 matrix
\begin{align*}
 \bm \tau_x = \left( 
   \begin{array}{cc}
       \bm 0  & \bm 1 \\
       \bm 1  & \bm 0  
   \end{array} \right).
\end{align*}
For brevity, we have dropped the arguments $\bm \varepsilon$ and $\phi$, but it is implied that $a \equiv a(\varepsilon)$ and $S_n \equiv S_n(\phi)$. 

(ii) The scattering matrix~(\ref{S_equal_gaps}) may be simplified to:
\begin{align}
  \bm S = - \frac{1}{a}\bm \tau_x + \frac{1-a^2}{a} [\bm I - a \bm S_n \bm\tau_x]^{-1} \bm \tau_x. \label{S_equal_gaps1}
\end{align}
An evident consequence of Eq.~(\ref{S_equal_gaps1}) is that the scattering matrix simplifies, $\bm S = -\bm \tau_x$, at the gap edge, i.e. at $\varepsilon = \Delta$, where $a = 1$. The $\phi$-dependence enters Eq.~(\ref{S_equal_gaps1}) only in the second term. Therefore, it is convenient to obtain the $\phi$-derivatives appearing in the trace of Eq.~(12) of the main text as
\begin{align*}
    &{\rm Tr}\left\{ \partial_{\phi}\bm S^\dagger\,\partial_{\phi}\bm S \right\} \\
    &= \frac{(1-a^2)^2}{a^2}{\rm Tr}\left\{\partial_{\phi} [\bm I-a\,\bm \tau_x \bm S_n^\dagger]^{-1}\,\partial_{\phi}[\bm I-a\, \bm S_n\bm \tau_x]^{-1}  \right\},
\end{align*}
where we permuted under the trace to eliminate additional $\bm \tau_x$ matrices and also used that $(\bm X^\dagger)^{-1} = (\bm X^{-1})^\dagger$ for invertible matrices ${\bm X}$.

(iii) Using that $\partial \bm X^{-1} = - \bm X^{-1} \partial \bm X \bm X^{-1}$ for invertible matrices ${\bm X}$, we may evaluate the $\phi$-derivatives under the trace as follows:
\begin{align}
    &{\rm Tr}\left\{ \partial_{\phi}\bm S^\dagger\,\partial_{\phi}\bm S \right\} \nonumber\\
    &= \frac{(1-a^2)^2}{a^2}\,{\rm Tr}\left\{[\bm I-a\,\bm \tau_x \bm S_n^\dagger]^{-1}[-\partial_{\phi}\,a\,\bm \tau_x \bm S_n^\dagger] [\bm I-a\,\bm \tau_x \bm S_n^\dagger]^{-1} \right. \nonumber\\ 
    & \qquad\qquad\left.\times [\bm I-a\, \bm S_n\bm \tau_x]^{-1}\,[-\partial_{\phi}\,a\, \bm S_n\bm \tau_x]\,[\bm I-a\, \bm S_n\bm \tau_x]^{-1}  \right\} \nonumber\\
    &= (1-a^2)^2\,{\rm Tr}\left\{[(\bm I-a\,\bm \tau_x \bm S_n^\dagger)(\bm I-a\, \bm S_n\bm \tau_x)]^{-1}\,\bm \tau_x\,\partial_{\phi}\bm S_n^\dagger\,  \right. \nonumber\\ 
    & \qquad\qquad\left.\times [(\bm I-a\, \bm S_n\bm \tau_x)(\bm I-a\,\bm \tau_x \bm S_n^\dagger)]^{-1}\,\partial_{\phi} \bm S_n\,\bm \tau_x\,  \right\} \nonumber\\
    &= (1-a^2)^2\,{\rm Tr}\left\{[\bm I(1+a^2)-a\,(\bm \tau_x \bm S_n^\dagger + \bm S_n\bm \tau_x)]^{-1}\,\bm \tau_x\,\partial_{\phi}\bm S_n^\dagger\,  \right.  \nonumber\\ 
    & \qquad\qquad\left.\times [\bm I(1+a^2)-a\,(\bm \tau_x \bm S_n^\dagger + \bm S_n\bm \tau_x)]^{-1}\,\partial_{\phi} \bm S_n\,\bm \tau_x\,  \right\}, \label{trace_ap1}
\end{align}
where in the penultimate line we used $X^{-1}Y^{-1} = [Y X]^{-1}$ and evaluated the corresponding products in Eq.~(\ref{trace_ap1}).

(iv) It is possible to check that 
\begin{align*}
    &[\bm I(1+a^2)-a\,(\bm \tau_x \bm S_n^\dagger + \bm S_n\bm \tau_x)] \\
    &\qquad\times[\bm I(1+a^2)+a\,(\bm \tau_x \bm S_n^\dagger + \bm S_n\bm \tau_x)]  = D \,\bm I ,
\end{align*}
by substituting the explicit expression for $\bm S_n$. Here $D = 1 + a^4 - 2a^2 (r^2+t^2\cos 2\phi)$ is (an energy-dependent) scalar. Thus, the inverse matrices appearing in Eq.~(\ref{trace_ap1}) may be written explicitly as
\begin{align*}
    &[\bm I(1+a^2)-a\,(\bm \tau_x \bm S_n^\dagger + \bm S_n\bm \tau_x)]^{-1} = \\
    &\qquad\frac{1}{D}[\bm I(1+a^2)+a\,(\bm \tau_x \bm S_n^\dagger + \bm S_n\bm \tau_x)].
\end{align*}
Using this relation in Eq.~(\ref{trace_ap1}), we obtain
\begin{align}
    &{\rm Tr}\left\{ \partial_{\phi}\bm S^\dagger\,\partial_{\phi}\bm S \right\} \nonumber\\
    &= \frac{(1-a^2)^2}{D^2}\,{\rm Tr}\left\{[\bm I(1+a^2)+a\,(\bm \tau_x \bm S_n^\dagger + \bm S_n\bm \tau_x)]\,\bm \tau_x\,\partial_{\phi}\bm S_n^\dagger\,  \right.  \nonumber\\ 
    & \qquad\qquad\qquad\left.\times [\bm I(1+a^2)+a\,(\bm \tau_x \bm S_n^\dagger + \bm S_n\bm \tau_x)]\,\partial_{\phi} \bm S_n\,\bm \tau_x \right\}. \nonumber
\end{align}

(v) We expand the matrix appearing in the latter equation in powers of $\bm \tau_x$. Only the even-power terms contribute to the trace, so we may write
\begin{align*}
    &{\rm Tr}\left\{ \partial_{\phi}\bm S^\dagger\,\partial_{\phi}\bm S \right\} \nonumber\\
    &= \frac{(1-a^2)^2}{D^2}\,{\rm Tr}\left\{(1+a^2)^2\,\bm \tau_x\,\partial_{\phi}\bm S_n^\dagger\,\partial_{\phi} \bm S_n\,\bm \tau_x\,  \right. \nonumber\\
    &  \left.+a^2\,(\bm \tau_x \bm S_n^\dagger + \bm S_n\bm \tau_x)\,\bm \tau_x\,\partial_{\phi}\bm S_n^\dagger\, (\bm \tau_x \bm S_n^\dagger + \bm S_n\bm \tau_x)\,\partial_{\phi} \bm S_n\,\bm \tau_x  \right\}. \nonumber    
\end{align*}
By using the expression for $\bm S_n$ in terms of the matrices Eq.~(\ref{S_N}), the trace may be evaluated explicitly,
\begin{align*}
    & {\rm Tr}\left\{\bm \tau_x\,\partial_{\phi}\bm S_n^\dagger\,\partial_{\phi} \bm S_n\,\bm \tau_x\,  \right\} = 4 t^2 \nonumber\\
    & {\rm Tr}\left\{(\bm \tau_x \bm S_n^\dagger + \bm S_n\bm \tau_x)\,\bm \tau_x\,\partial_{\phi}\bm S_n^\dagger\, (\bm \tau_x \bm S_n^\dagger + \bm S_n\bm \tau_x)\,\partial_{\phi} \bm S_n\,\bm \tau_x\,  \right\}   \nonumber    \\
    & \qquad = 8 t^2 [t^2 + (2-t^2) \cos 2\phi].
\end{align*}
Finally, collecting all terms together,we obtain:
\begin{align*}
    &{\rm Tr}\left\{ \partial_{\phi}\bm S^\dagger\,\partial_{\phi}\bm S \right\} \nonumber\\
    &= \frac{4t^2(1-a^2)^2 \left[ (1+a^2)^2 + 2 a^2 t^2 + 2a^2(2-t^2) \cos 2\phi \right]}
    {\left[(1-a^2)^2+2a^2t^2(1-\cos 2\phi)\right]^2}.\,  
\end{align*}

(vi) Integration of the above expression over $\phi$ reproduces the integrand in Eq.~(15) of the main text:
\begin{align*}
    &\int \frac{d\phi}{2\pi}\,{\rm Tr}\left\{ \partial_{\phi}\bm S^\dagger\,\partial_{\phi}\bm S \right\} \nonumber\\
    & \qquad = \frac{4 t^2(1+a^2)^2}{(1-a^2)\sqrt{(1-a^2)^2+4a^2t^2}} \\
    & \qquad = \frac{4 t^2\varepsilon^2}{\sqrt{(\varepsilon^2-\Delta^2)(\varepsilon^2-\Delta^2(1-t^2))}}. 
\end{align*}

\section{Conductance of the NS junction ($\Delta_1= 0$, $\Delta_2 = \Delta$).}
Below, we provide details of derivation of the conductance of NS junction. Our goal here is to show how the known results~\cite{BTK1982} come out from Eq.~(12) of the main text.

(i) We consider the case where the left lead is normal, i.e. $\Delta_1 = 0$, whereas the right lead is superconducting, i.e. $\Delta_2 = \Delta$. This induces the following Andreev scattering amplitudes $a_1 = 0$ and $a_2 = (\varepsilon-\sqrt{\varepsilon^2-\Delta^2})/\Delta \equiv a$. We also introduce $b = \sqrt{1-a^2}$, which has a meaning of transmission amplitude through a clean NS boundary. Plugging them in Eq.~(\ref{S_SNS}), one may obtain the 4-by-4 scattering matrix
\begin{align}
    & \bm S \equiv \left(\begin{array}{cccc}
        r^{ee}_{11} & t^{ee}_{12} & r^{eh}_{11} & t^{eh}_{12} \\ 
        t^{ee}_{21} & r^{ee}_{22} & t^{eh}_{21} & r^{eh}_{22} \\
        r^{he}_{11} & t^{he}_{12} & r^{hh}_{11} & t^{hh}_{12} \\
        t^{he}_{21} & r^{he}_{22} & t^{hh}_{21} & r^{hh}_{22} 
    \end{array}\right) \label{S_NS} \\ 
    &= \frac{1}{D}\left(
\begin{array}{cccc}
 b^2r & ibt\,e^{-i\phi} & at^2\, e^{-2i\phi}  & iabrt\, e^{-i \phi }  \\
 ibt\, e^{i\phi}  & b^2r & -iabrt\, e^{-i\phi}  & -at^2 \\
 at^2 \,e^{2i\phi}  & -iabrt\, e^{i\phi}& b^2r & -ibt\, e^{i\phi}  \\
 iabrt\,e^{i\phi}& -at^2 & -ibt\, e^{-i\phi}  & b^2r 
\end{array} \nonumber
\right),
\end{align}
where $D = 1-a^2r^2$. In the first line of Eq.~(\ref{S_NS}), we gave an explicit representation of the matrix elements of the scattering matrix in terms of the scattering amplitudes, i.e. $t^{ee}_{21},\,t^{he}_{21}$, etc. Here the top indices label the particle-hole branches, whereas the bottom indices label the leads. For example, $t^{he}_{21}$ represents a process of a particle-like quasiparticle incident from the left (1) lead which scattering into a hole-like state in the right (2) lead. 

(ii) {\it Quasiparticles with low energies $\varepsilon<\Delta$.} Consideration of the contribution to $G$ of the excitations with energy $\varepsilon<\Delta$ impinging on the interface from the normal lead is especially simple and insightful.
In this case, the function $a(\varepsilon)$ becomes complex, $a(\varepsilon) = (\varepsilon-i\sqrt{\Delta^2-\varepsilon^2})/\Delta$. At energies $\varepsilon<\Delta$ excitations reside only in the left (normal) lead, so that the scattering matrix~(\ref{S_NS}) reduces to 2-by-2
\begin{align}
    & \bm S \equiv \left(\begin{array}{cc}
        r^{ee}_{11} & r^{eh}_{11} \\ 
        r^{he}_{11} & r^{hh}_{11} 
    \end{array}\right) =  \frac{1}{1-a^2r^2}\left(
\begin{array}{cc}
 b^2r & at^2\, e^{-2i\phi} \\
 at^2 \,e^{2i\phi}  & b^2r 
\end{array} 
\right).\label{S_NS_supgap}
\end{align}
The only $\phi$-dependence here comes from Andreev reflection processes encoded in the exponential prefactors $\propto e^{2i\phi}$ of the off-diagonal elements.
Using Eq.~(\ref{S_NS_supgap}) it is straightforward to evaluate the $\phi$-integral in Eq.~(12):
\begin{align}
    \int\limits_{0}^{2\pi} \frac{d\phi}{2\pi} \,   {\rm Tr}\left\{ \partial_{\phi}\bm S^\dagger(\phi,\varepsilon)\,\partial_{\phi}\bm S(\phi,\varepsilon) \right\} = 4 |r^{eh}_{11}|^2 + 4 |r^{he}_{11}|^2. \nonumber
\end{align}
Unitarity of the S-matrix allows us to re-write the latter equation as 
\begin{align}
   & \int\limits_{0}^{2\pi} \frac{d\phi}{2\pi} \,   {\rm Tr}\left\{ \partial_{\phi}\bm S^\dagger(\phi,\varepsilon)\,\partial_{\phi}\bm S(\phi,\varepsilon) \right\} \nonumber \\
   & \qquad = 2(1- |r^{ee}_{11}|^2 + |r^{eh}_{11}|^2)+2(1- |r^{hh}_{11}|^2 + |r^{he}_{11}|^2).  \label{subgap}
\end{align}

(iii) {\it Quasiparticles with energies $\varepsilon>\Delta$}. Here the full 4-by-4 matrix~(\ref{S_NS}) must be considered. In addition to the Andreev processes, it also contains single-particle transmission amplitudes decorated by factors $e^{i\phi}$.
Evaluation of the proper trace is straightforward,
\begin{align}
    &\int\limits_{0}^{2\pi} \frac{d\phi}{2\pi} \,   {\rm Tr}\left\{ \partial_{\phi}\bm S^\dagger(\phi,\varepsilon)\,\partial_{\phi}\bm S(\phi,\varepsilon) \right\}  \nonumber \\
    & \qquad = (4|r^{he}_{11}|^2 + |t^{ee}_{21}|^2 + |t^{he}_{21}|^2 + |t^{ee}_{12}|^2 + |t^{he}_{12}|^2)   \nonumber \\
    &\qquad\qquad + (4|r^{eh}_{11}|^2 + |t^{hh}_{21}|^2 + |t^{eh}_{21}|^2 + |t^{hh}_{12}|^2 + |t^{eh}_{12}|^2).  \nonumber 
\end{align}
Noting that  $|t^{bb'}_{12}|=|t^{bb'}_{21}|$ for each quasiparticle branch and using the unitarity of the scattering matrix, it can be further simplified,
\begin{align}
   & \int\limits_{0}^{2\pi} \frac{d\phi}{2\pi} \,   {\rm Tr}\left\{ \partial_{\phi}\bm S^\dagger(\phi,\varepsilon)\,\partial_{\phi}\bm S(\phi,\varepsilon) \right\} \nonumber \\
   & \qquad = 2(1- |r^{ee}_{11}|^2 + |r^{eh}_{11}|^2)+2(1- |r^{hh}_{11}|^2 + |r^{he}_{11}|^2).  \label{supragap}
\end{align}

(iv) Given the identical form of Eqs.~(\ref{subgap}) and (\ref{supragap}),
we may write the conductance using Eq.~(12) as
\begin{eqnarray}
    G_{NS} = \frac{2e^2}{h} \int_0^{\infty} d\varepsilon\,[&-&\partial_\varepsilon f(\varepsilon)]\left[(1 -|r^{ee}|^2+|r^{he}|^2)\right.
    \nonumber\\
    &+&(1-|r^{hh}|^2+\left.|r^{eh}|^2)\right]\,.
    \label{GNS_ap}
\end{eqnarray}
The terms in the first and second parentheses represent, respectively, the particle-like and hole-like contributions to the conductance. Equation~(\ref{GNS_ap})
agrees with the well-known expression for NS junctions \cite{BTK1982}. The concrete expression in terms of the transmission coefficient $\tau$ may also be evaluated after some algebra
\begin{align}
    \frac{G_{NS}}{G_n} &= \int_0^\Delta d\varepsilon\,\frac{4\tau \Delta^2}{\Delta^2(2-\tau)^2-4\varepsilon^2(1-\tau)}[-\partial_\varepsilon f(\varepsilon)] \nonumber \\
    &  + \int_\Delta^\infty d\varepsilon\,\frac{4\,\varepsilon}{\tau\varepsilon+(2-\tau)\sqrt{\varepsilon^2-\Delta^2}}[-\partial_\varepsilon f(\varepsilon)]. \label{conductance_NS_ap}
\end{align}

\section{Asymptotic behavior of the superconducting point contact conductance at low $T \ll \Delta$ and high $T\gg \Delta$ temperatures.} \label{sec:conductance_SNS_asymptotic}
Below, we provide details of finding asymptotic behavior of the formula for SPC conductance,
with reference to Eq.~(15) of the main text. It is convenient to switch there to a dimensionless integration variable $x$ defined as $\varepsilon/T = x$, so that integral becomes
\begin{align}
    \frac{G_{SPC}}{G_n} = \int_\alpha^\infty dx \, \frac{2\,x^2}{\sqrt{(x^2-\alpha^2)(x^2-\alpha^2(1-\tau))}}[-f'(x)], \label{intX}
\end{align}
where $\alpha = \Delta/T$ and $f(x) = (e^x+1)^{-1}$.

(i) {\it Low temperature ($\alpha\gg 1$) asymptote.} Here the Fermi function may be approximated with the Boltzmann distribution, $f(x) \approx e^{-x}$. It is convenient to  further shift the integration variable, $x \to x+\alpha$,
\begin{align*}
        &\frac{G_{SPC}}{G_n} \approx \\ &\int_0^\infty dx \, \frac{2\,(x+\alpha)^2}{\sqrt{[(x+\alpha)^2-\alpha^2][(x+\alpha)^2-\alpha^2(1-\tau)]}}e^{-x-\alpha}.
\end{align*}
The terms $x$ can be neglected with respect to large $\alpha$ in the appropriate places of the integrand, which renders
\begin{align*}
        &\frac{G_{SPC}}{G_n} \approx \\ &\alpha \sqrt{\frac{2}{1+\sqrt{1-\tau}}}\,e^{-\alpha}\int_0^\infty dx \, \frac{e^{-x}}{\sqrt{x[x+\alpha(1-\sqrt{1-\tau})]}}.
\end{align*}
The integral may be recognized as the Bessel function $K_0(\lambda)=\int_1^\infty dx\,\frac{e^{-x\lambda}}{\sqrt{x^2-1}}$, and, thus, the conductance becomes 
\begin{align*}
        &\frac{G_{SPC}}{G_n} \approx \alpha\, \sqrt{\frac{2}{1+\sqrt{1-\tau}}}\,e^{-\frac{\alpha}2(1+\sqrt{1-\tau})}\,K_0\left[\frac{\alpha}{2}(1-\sqrt{1-\tau})\right].
\end{align*}
The asymptotes of the Bessel function here 
are $K_0(\lambda) \approx \sqrt{\frac{\pi}{2\lambda}}\,e^{-\lambda}$ for $\lambda\gg 1$ and $K_0(\lambda) \approx \ln \left( \frac{2 e^{-\gamma}}{\lambda}\right)$  for $\lambda \ll 1$.

\begin{widetext}
{\it High-temperature ($\alpha  \ll 1$) asymptote.} In this limit, it is convenient to single out the trivial term in Eq. (\ref{intX}),
\begin{align*}
    \frac{G_{SPC}}{G_n} =2\int_\alpha^\infty dx \, [-f'(x)] + 2\int_\alpha^\infty dx \, \left[\frac{x^2}{\sqrt{(x^2-\alpha^2)(x^2-\alpha^2(1-\tau))}}-1\right][-f'(x)], 
\end{align*}
at the expense of introducing $-1$ in the integrand of the second integral. The first integral is evaluated yielding $2f(\alpha)$. In the second integral, we bring the terms in the square brackets to the same denominator, multiply the numerator and denominator of the resulting fraction by the conjugate expression, and further switch to a new integration variable $x \to \alpha x$. Thus, we obtain
\begin{align*}
    \frac{G_{SPC}}{G_n} =2f(\alpha) + 2\alpha \int_1^\infty dx \, \left[\frac{(x^2-1)(2-\tau)+1}{\sqrt{(x^2-1)(x^2-1+\tau)}(x^2+\sqrt{(x^2-1)(x^2-1+\tau)})}\right][-f'(\alpha x)], 
\end{align*}
Note that the expression in the square brackets of the integrand behaves as $1/x^2$ at large $x$, so the integral converges well. Thus, one may replace $-f'(\alpha x) \approx 1/4$ to the leading order at small $\alpha$. That together with an expansion $f(\alpha) =1/2 - \alpha/4$ gives an asymptotic approximation of the conductance at $\alpha = \Delta/T \ll 1$
\begin{align}
    \frac{G_{SPC}}{G_n} \approx 1 +\alpha\, k(\tau),
\end{align}
where we introduced a $\tau$-dependent function 
\begin{align}
    k(\tau) = -\frac{1}{2}+\frac{1}{2}\int_1^\infty dx\, \frac{(x^2-1)(2-\tau)+1}{\sqrt{(x^2-1)(x^2-1+\tau)}(x^2+\sqrt{(x^2-1)(x^2-1+\tau)})}.
\end{align}
The function $k(\tau)$ is positive on the interval $1>\tau>0$; it is logarithmically large $k(\tau) \approx -\frac{1}{4}\ln(\tau)$ at small $\tau\ll 1$ and vanishes $k(1) = 0$ at perfect transmission $\tau = 1$. 
\end{widetext}

\section{Analysis of the series in $N$ in Eq. (10).} \label{sec:TN_ap}
In the main part of the Letter, we obtained a representation of the absorbed power $\mathcal P$ via a series in $N$, see Eq.~(10) of the main text. The integer $N$ stands for the number of absorbed/released energy quanta. The purpose of this section is to analyze the convergence of that series in $N$. 

(i) For definiteness, we focus on a specific matrix element of the elastic scattering matrix $t^{he}(\phi)$ 
of a short channel connecting two superconducting leads. It has a  simple form, \cite{thermopower2019} 
\begin{align*}
    t^{he}(\phi) = - r \frac{\xi}{t\Delta}\,\frac{\sin\phi}{(\xi/t\Delta)^2+\sin^2\phi},
\end{align*}
where $\xi = \sqrt{\varepsilon^2-\Delta^2}$. The scattering amplitude $t^{he}(\phi)$ contributes to transport even in the tunnelling ($t \ll 1$) regime. Analysis of other amplitudes can be performed in a similar way.

(ii) The function $t^{he}(\phi)$ is periodic in $\phi$ and has only odd harmonics $t^{he}(\phi) = \sum_N \widetilde{t^{he}}(N) e^{i\phi N}$, where $N = 2n+1$.  One may evaluate them:
\begin{align}
    \widetilde{t^{he}}(N) = -\frac{r}{2i} \frac{\xi}{t\Delta} \frac{1}{\sqrt{1+\left(\frac{\xi}{t\Delta}\right)^2}\left[\left(\frac{\xi}{t\Delta}\right) + \sqrt{1+\left(\frac{\xi}{t\Delta}\right)^2}\right]^N}. 
    \label{theN}
\end{align}
Recall that this Fourier harmonic also corresponds to the scattering amplitude with absorption of $N$ photons according to Eq.~(7) of the main text. 

(iii) Next, we seek to average the probability of that process over energy in the Gibbs ensemble,
\begin{align*}
    P_N = \int_\Delta^\infty d\varepsilon\, \left| \widetilde{t^{he}}(N)\right|^2 [-\partial_\varepsilon f(\varepsilon)].
\end{align*}
To be specific, it corresponds to the integral $(2\pi)^2\int d\varepsilon\, \rho_1(\varepsilon)\rho_2(\varepsilon) \left| \mathcal{\tilde T}_{1\varepsilon\,\,2\varepsilon}(N,0)\right|^2 \,[-\partial_{\varepsilon}f(\varepsilon)]$ in the notations of Eq.~(10). For simplicity, we focus on a case of low temperatures $T\ll \Delta$, where the Fermi function simplifies $[-\partial_\varepsilon f(\varepsilon)] = \frac1T e^{-\varepsilon/T}$. In addition, one may expand $\varepsilon \approx \Delta + \xi^2/2\Delta$ and switch to the integration variable $\xi$. So, the integral becomes
\begin{align*}
 P_N = \frac{1}{T}e^{-\Delta/T}\int_0^\infty d\xi \, \frac{\xi}{\Delta}\, |\widetilde{t^{he}}(N)|^2\, e^{-\xi^2/2\Delta T}. 
\end{align*}
Note that $\xi\sim t\Delta$ is an effective energy scale at which $\widetilde{t_{he}}(N)$ changes, whereas the exponential term in the integrand changes at the scale $\xi \sim \sqrt{\Delta T}$. In order to compare the two scales, it is instructive to switch to a dimensionless integration variable $x = \xi/t\Delta$. So, we substitute Eq.~(\ref{theN}) in the last equation and obtain 
\begin{align}
    P_N =& \frac{\tau(1-\tau)\Delta}{T}\,e^{-\Delta/T}\, \nonumber\\
    &\quad \times\int_0^\infty dx\, x^3 \frac{e^{-x^2 \Delta \tau/2T}}{(1+x^2)(x+\sqrt{1+x^2})^{2N}}, \label{PN}
\end{align}
where $\tau = t^2$.

(iv) {\it Limit of small $\tau \Delta/T \ll 1$.}  In this limit, the exponential factor $e^{-x^2 \Delta \tau/2T}$ may be neglected for $N>1$ because the integrand $\propto 1/x^{2N-1}$ converges well at $N>1$. In order to estimate the behavior of $P_N$ at large $N$, we notice that small $x\ll 1$ contribute most to the integral, so one may approximate the denominator of the integrand as $(1+x^2)(x+\sqrt{1+x^2})^{2N} \approx (1+x)^{2N} \approx e^{2xN}$. It is, then, straightforward to find the asymptotics  $P_N \propto 1/N^4$ at large $N$.

However at $N = 1$, it is crucial to retain the exponential term $e^{-x^2 \Delta \tau/2T}$, which cuts off the logarithmic divergence and produces $\propto \ln (T/\Delta\tau)$ after integration. It is interesting to note that it is the processes with the absorption or release of  $N=1$ energy quanta that contribute most to transport at small $\tau$. 

(iv) {\it Limit of large $\tau\Delta/T \gg 1$.} 
Because the integral converges at small $x$, we may also approximate the denominator $(1+x^2)(x+\sqrt{1+x^2})^{2N} \approx e^{2xN}$. So, one may rewrite the integral in Eq.~(\ref{PN}),
\begin{align*}
    P_N \propto\int_0^\infty dx\, x^3\, e^{-2Nx}\,e^{-x^2\Delta\tau/2T}.
\end{align*}
The competition of the two exponential factors determines the evolution of $P_N$ with $N$. For $N \ll N^\ast \sim \sqrt{\tau\Delta/T}$, the last exponential term dominates, so the integral depends weakly on $N$ producing a plateau in $P_N$. For $N\gg N^\ast \sim \sqrt{\tau \Delta/T}$, the first exponential term dominates, producing $P_N \propto 1/N^4$. 

We conclude that the processes with the absorption of a large number of energy quanta (up to $N^\ast \sim \sqrt{\tau\Delta/T}$) are important at any $\tau$, as long as the condition $\tau\Delta/T \gg 1$ is satisfied.
\section{Expansion of $I(\mathcal V)$ in powers of $\mathcal V$ in the tunnelling limit.} \label{sec:IV_expansion}
In the tunnelling limit $\tau\ll 1$, the full $I(\mathcal V)$ dependence is known,~\cite{Larkin1967,TinkhamBook}
\begin{align}
I(\mathcal V) =& \frac{G_n}{e} \int_{\Delta_2}^\infty d\varepsilon\,\left\{\rho_1(\varepsilon+e\mathcal V)\,\rho_2(\varepsilon) \left[f(\varepsilon)-f(\varepsilon+e\mathcal V)\right]\right. \nonumber \\
 & \qquad\,\,\left. -\rho_1(\varepsilon-e\mathcal V)\,\rho_2(\varepsilon)  \left[f(\varepsilon)-f(\varepsilon-e\mathcal V)\right] \right\}. \label{tunneling_current}
\end{align}
Here $\rho_{1,2}(\varepsilon) = \varepsilon/\sqrt{\varepsilon^2-\Delta_{1,2}^2}$ are the normalized densities of states in the two superconducting leads. For concreteness, we assume that the gaps are not equal and $\Delta_2 > \Delta_1$. Current is an odd function of $\mathcal V$, so expansion of $I$ in $\mathcal V$ has only odd terms, $I(\mathcal V) = I_1(\mathcal V) + I_3(\mathcal V) + \mathcal O(\mathcal V^5)$ with $I_n \propto \mathcal V^n$. We find them by expanding~(\ref{tunneling_current}) in ${\cal V}$,
\begin{align}
    &I_1(\mathcal V) = -2G_n \mathcal V \int^\infty_{\Delta_2} d\varepsilon\,\rho_1(\varepsilon) \rho_2(\varepsilon)\,\partial_\varepsilon  f(\varepsilon), \label{I1V}\\
    &I_3(\mathcal V) = -2G_n e^2\mathcal V^3\,\int_{\Delta_2}^{\infty}d\varepsilon\,\rho_2(\varepsilon) \left[\frac{\rho_1(\varepsilon)\,\partial^3_\varepsilon f(\varepsilon)}{3!}\right. \nonumber \\
    & \qquad\qquad\left. + \frac{\partial_\varepsilon \rho_1(\varepsilon) \,\partial^2_\varepsilon f(\varepsilon)}{2} + \frac{\partial^2_\varepsilon \rho_1(\varepsilon)\, \partial_\varepsilon f(\varepsilon)}{2} \right]. \label{I3V}
\end{align} 
In order to compare the relative importance of the linear $I_1(\mathcal V)$ and non-linear $I_3(\mathcal V)$ currents we evaluate them in the limit of small temperatures $T \ll \Delta_1 \sim \Delta_2$. We find that the ratio of the currents scales as $\frac{I_3}{I_1} \propto \left(\frac{e\mathcal V}{T}\right)^2$ and $\frac{I_3}{I_1} \propto \left(\frac{e\mathcal V}{\Delta_2-\Delta_1}\right)^2 \frac{1}{\ln[T/(\Delta_2-\Delta_1)]}$ in the regimes $T \ll (\Delta_2-\Delta_1)$ and $T \gg (\Delta_2 - \Delta_1)$, respectively. This allows us to conclude that the non-linear term $I_3$ may be neglected as long as $e\mathcal V \ll \min(T,|\Delta_2-\Delta_1|)$ is the smallest energy scale.

\section{Dissipative current carried by the Andreev levels in the presence of relaxation.} \label{sec:AndreevDissipative}

(i) A short SPC supports Andreev levels with energies $\pm \varepsilon_{A}(\phi)$, where $\varepsilon_A(\phi) = \Delta\sqrt{1-\tau\sin\phi^2}$. Note the conventionally-defined Josephson phase difference is $\varphi = 2\phi$. At thermodynamic equilibrium, the electric current carried by the Andreev levels is given by the Josephson relation
\begin{align}
    I_A(\phi) = -\frac{e}{\hbar}\,\partial_{\phi}\varepsilon_A(\phi)\, n_{eq}[\varepsilon_A(\phi)],
\end{align}
where $n_{eq}[\varepsilon_A] = f[-\varepsilon_A]-f[\varepsilon_A] = \tanh [\varepsilon_A/2T]$ is the difference of the fermionic occupations of the Andreev levels with negative $-\varepsilon_A$ and positive $\varepsilon_A$ energies.

(ii) Under the applied voltage bias $\mathcal V$, the phase winds, $\phi(t) = \Omega \, t$, with frequency $\Omega = e\mathcal V/\hbar$. This results in an ac Josephson effect, where the Josephson current $I_A(\phi(t))$ oscillates with frequency $2\,\Omega$. If averaged over the period of oscillations, the net current vanishes. However, relaxation may result in the non-zero net current due to the Debye mechanism \cite{Debye1912}. In order to describe it, we follow Ref.~[\onlinecite{AverinAdiabatic1996}] and introduce a phenomenological relaxation rate $\gamma$ of the occupation function
\begin{align}
    \frac{dn}{dt} + \gamma \, n = \gamma\, n_{eq}[\varepsilon_A(\phi(t))].
\end{align}

(iii) The solution of this equation may be presented in a general form 
\begin{align}
    n(t) = \gamma \int_{-\infty}^t dt'\, e^{-\gamma(t-t')}\, n_{eq}[\varepsilon_A(\phi(t'))].
\end{align}
This allows us to write the current $I_A(t) =-\frac{e}{\hbar}\,\partial_{\phi}\varepsilon_A(\phi(t))\, n(t)$ and average it over the period of oscillations
\begin{align*}
    &\langle I_A \rangle =\frac{\Omega}{\pi}\,\int_{0}^{\frac{\pi}{\Omega}}\,dt\,I_A(t)= -\frac{\Omega \,e}{\pi\,\hbar}\int_{0}^{\frac{\pi}{\Omega}}\,dt\,\partial_{\phi}\varepsilon_A(\phi(t)) \\
    &\qquad\times \gamma \int_{-\infty}^t dt'\, e^{-\gamma(t-t')}\, n_{eq}[\varepsilon_A(\phi(t'))].
\end{align*}
We change integration variables $t \to \phi/\Omega$ and $t' \to (\phi-\phi')/\Omega$, exchange the order of integration, and further massage it to the form 
\begin{align}
    &\langle I_A\rangle = \label{average_Josephson} \\
    &  -\frac{e}{\pi \hbar}\frac{\gamma}{\Omega}\,\int_{0}^{\infty}\,d\phi'\,e^{-\frac{\gamma}{\Omega} \phi'}\int_{0}^\pi \, d\phi\,\, \partial_{\phi}\,\varepsilon_A(\phi+\phi')\,    n_{eq}[\varepsilon_A(\phi)]. \nonumber
\end{align}

(iv) {\it Limit of small $\tau$, such that $\frac{\tau\Delta}{T} \ll 1$.} Then, the energy of the Andreev level can be approximated as $\varepsilon_A(\phi) \approx \Delta - \frac{\Delta \tau}{2}\sin^2 \phi$, so $n_{eq}[\varepsilon_A(\phi) ] \approx n_{eq}[\Delta] - \frac{\tau\Delta}{2}\sin^2\phi \,\, \partial_\varepsilon n_{eq}[\Delta]$. We plug these expansions in Eq.~(\ref{average_Josephson}) and obtain
\begin{align}
    &\langle I_A\rangle =   \frac{e}{\hbar}\, \frac{\tau^2\Delta^2}{8}\,\frac{\gamma\,\Omega}{\gamma^2+4\Omega^2}\,    \partial_\varepsilon n_{eq}[\Delta], \nonumber
\end{align}
where recall that $\Omega = e\mathcal V/\hbar$.

(v) {\it Limit of low temperatures $T/\Delta \ll 1$ and intermediate $\tau$ (i.e. $\tau \sim 1-\tau$).} The main contribution to the dissipation comes from the vicinity of $\phi = \pi/2$ where the energy of Andreev level reaches minimum, and at low temperatures ($\Delta\sqrt{1-\tau}/T\gg 1$) the occupation factor can be approximated as
\begin{align*}
    &n[\varepsilon_A(\phi)] \approx \\ &\,\,1-2\,\exp\left[-\frac{\Delta}{T}\sqrt{1-\tau}-\frac{\Delta}{T}\frac{\tau}{2\sqrt{1-\tau}}(\phi-\pi/2)^2\right]
.
\end{align*}
Because of the large factor $\Delta/T$ in the exponent, the integral over $\phi$ in Eq.~(\ref{average_Josephson}) converges fast. After performing integration and simplification one obtains
\begin{align}
    &\langle I_A \rangle = \frac{e\Delta}{\hbar}\,\frac{\gamma}{\Omega}\,\sqrt{\frac{T}{\Delta}}\,\exp\left(-\frac{\Delta}{T}\sqrt{1-\tau}\right)\, M(\gamma/\Omega,\tau), \nonumber
\end{align}
where $M(\gamma/\Omega,\tau)$ is a crossover ``memory'' function
\begin{align*}
    &M(\gamma/\Omega,\tau) =  \frac{\sqrt{(2/\pi)\tau\sqrt{1-\tau}}}{1-e^{-\gamma\pi/\Omega}}
    \\
    & \qquad\qquad \times \int_0^{\pi/2}\, d\phi\,\left[e^{-\frac{\gamma}{\Omega}\phi}-e^{-\frac{\gamma}{\Omega}(\pi-\phi)}\right]\frac{\sin 2\phi}{\sqrt{1-\tau\cos^2\phi}}.
\end{align*}
In the limit $\gamma/\Omega\ll 1$, it has a finite value, $M(0,\tau)\sim 1$ at intermediate $\tau$, and $M(0,\tau)\sim \sqrt{\tau}$ at $\tau\to 0$. Dispensing with the weak dependence on $\tau$, we conclude that
\begin{align}
    &\langle I_A \rangle \sim \frac{e\Delta}{\hbar}\,\frac{\gamma}{\Omega}\,\sqrt{\frac{T}{\Delta}}\,\exp\left(-\frac{\Delta}{T}\sqrt{1-\tau}\right)
    \label{IAfinal}
\end{align}
is proportional to the small parameter $\gamma/\Omega$. In the opposite limit $\gamma/\Omega\gg 1$ the asymptote of $M$ reads $M(\gamma/\Omega,\tau) \approx \sqrt{\frac{8\tau}{\pi(1-\tau)}}\,\left(\frac{\Omega}{\gamma}\right)^2$, so that the $\gamma/\Omega$ factor in Eq.~(\ref{IAfinal}) is replaced with its inverse.


\end{document}